\documentstyle[aps]{revtex}

\newcommand{\ket}[1]{|{#1}\rangle}
\newcommand{\bra}[1]{\langle{#1}|}
\newcommand{\kets}[2]{|{#1}\rangle_{{}_{\!\!{#2}}}}

\newcommand{\slb}[2]{{#1}^{({#2})}}
\newcommand{\cA}{{\cal A}}
\newcommand{\cC}{{\cal C}}
\newcommand{\cE}{{\cal E}}

\newcommand{\cH}{{\cal H}}
\newcommand{\cJ}{{\cal J}}
\newcommand{\cL}{{\cal L}}

\newcommand{\cS}{{\cal S}}
\newcommand{\cZ}{{\cal Z}}
\renewcommand{\tensor}{\otimes}
\newcommand{\trace}{\mbox{tr}}
\newcommand{\vspan}{\mbox{span}}

\newtheorem{Theorem}{Theorem}

\newcommand{\proof}{\paragraph*{Proof.}}
\newcommand{\proofof}[1]{\paragraph*{Proof of Thm.#1.}}
\newcommand{\qed}{\hspace*{\fill}\rule{2.5mm}{2.5mm}%
\vspace*{8pt}\par}

\newcommand{\phead}[1]{\par\noindent{\bf #1}}

\begin{document}
\twocolumn

\title{Theory of Quantum Error Correction for General Noise}

\author{Emanuel Knill$^1$, Raymond Laflamme$^1$, and Lorenza Viola$^2$
\thanks{E-mail addresses: {\tt knill@lanl.gov; laflamme@lanl.gov;}
{\tt vlorenza@mit.edu} } } 
\address{ $^1$ Newton Institute, Cambridge, United Kingdom, and \\
Los Alamos National Laboratory, MS B265, Los Alamos, New Mexico 87545 \\
$^2$ d'Arbeloff Laboratory for Information Systems and Technology, 
Department of Mechanical Engineering,  \\
Massachusetts Institute of Technology,  Cambridge, Massachusetts 02139 } 

\date{\today}
\maketitle

\begin{abstract}
Quantum error correction protects quantum information against
environmental noise. When using qubits, a measure of quality of a code
is the maximum number of errors that it is able to correct. We show
that a suitable notion of ``number of errors'' $e$ makes sense for any
system in the presence of arbitrary environmental interactions.  In
fact, the notion is directly related to the lowest order in time with
which uncorrectable errors are introduced, and this in turn is derived
from a grading of the algebra generated by the interaction operators.
As a result, $e$-error-correcting codes are effective at protecting
quantum information without requiring the usual assumptions of
independence and lack of correlation.  We prove the existence of large
codes for both quantum and classical information.  By viewing
error-correcting codes as subsystems, we relate codes to irreducible
representations of certain operator algebras and show that noiseless
subsystems are infinite-distance error-correcting codes.  An explicit
example involving collective interactions is discussed.
\end{abstract}

One of the main reasons for the robustness of quantum
computation~\cite{shor:qc1996a,aharonov:qc1996a,kitaev:qc1997a,knill:qc1998a}
is the ability to use quantum error-correcting
codes~\cite{shor:qc1995b,steane:qc1995a} to maintain information
stored in qubits (two-state particles) subject to environmental noise.
Quantum error-correcting codes are defined as subspaces of the qubits'
state space with the property that an initial state in this subspace
can be recovered if sufficiently few of the qubits have experienced
errors.  Provided the noise affecting different qubits is independent
and not too intense, any quantum information ({\em i.e.\/},~state)
stored in the subspace can then be regained with high fidelity.  This
view suffers from several disadvantages. Notably, it is not obvious
whether collective errors can also be corrected well, nor is it clear
in what sense the information is preserved {\em before} it is
recovered by correcting the errors. In addition, the present theory
does not directly lend itself to the application of similar ideas to
physical systems that are not canonically decomposable into qubits or
are subject to different interaction Hamiltonians.

In this Letter, we overcome the above inconveniences by introducing a
general description of arbitrary system-environment couplings in
terms of a {\em graded interaction algebra}. The degree of an operator
in this algebra both determines the temporal order and the extent to
which the operator can affect the system, independent of the internal
evolution of the environment.  In the case of qubits with independent
one-qubit interactions, this notion coincides with the usual concepts
of ``number of errors'' or ``error weight'' used in combinatorial
error analysis.  We find that the generalization of minimum distance
relates to error correction in the usual way and show that,
irrespective of the nature of the environmental noise, large codes
exist depending solely on the dimension of the linear space of errors of
a given order.

Using algebras to classify errors naturally leads to algebraic methods
for describing error-correcting codes. The basic idea is to revisit
the notion of error-correcting codes as ``abstract
particles''~\cite{knill:qc1995e}, which are associated with
irreducible representations ({\em irreps}) of operator algebras closed
under Hermitian conjugation ($^\dagger$-closed). Accordingly,
error-correcting codes can be viewed as {\em subsystems} ({\em
i.e.}~tensor factors of subspaces), which makes it clear where the protected
information resides and eliminates the need for error correction
except to make the information available in a standard form.  This
relates general error correction to a trivial example: Suppose that
errors affect all but the first qubit. Then information in the first
qubit is clearly safe, and the qubit can be regarded as {\em
noiseless}.  We show that {\em noiseless subsystems} not only are
identical to infinite-distance error-correcting codes, but they also
provide the most general method of noise-free information storage, 
thereby substantially extending the concept of noiseless
subspaces~\cite{zanardi:qc1998a,lidar:qc1998a}.

\phead{Systems and Noise.} Let $S$ be a quantum system with state
space $\cS$. $\cS$ is an $N$-dimensional Hilbert space.  $S$ interacts
with the environment $B$ via an interaction Hamiltonian $J$ which can
be written in the form
\begin{eqnarray}
J &=& \sum_i J_i\tensor B_i \:,
\label{eqn:j}
\end{eqnarray}
where the $B_i$'s are linearly independent environment operators. We
assume that the internal evolution of $B$ is removed from
$J$ by requiring that $\trace(J_i)=0$ for all $i$. The internal
evolution of $S$ is retained. If desirable, the latter can be 
absorbed into a rotating frame, at the expense of making the operators
$J_i$ explicitly time-dependent. This can be accommodated within the present
formalism through appropriate redefinitions of the relevant quantities. 
Our analysis only depends on the noise-inducing interaction (\ref{eqn:j}) 
through an overall {\em noise-strength} parameter given by
\begin{eqnarray}
\lambda &=& |J| \:,
\label{eqn:lambda}
\end{eqnarray}
where $|X|$ is the maximum eigenvalue of $\sqrt{X^\dagger X}$. The
above quantity can be infinite in situations involving
infinite-dimensional environments, {\em e.g.} the modes of an
electromagnetic field.  In such cases a redefinition of $\lambda$ is
necessary, based on additional information about the initial state and
the internal evolution of the environment. Important examples are
noise in the Markovian limit and discrete noise given by quantum
operations, which will be discussed later.

The second concept we introduce is the {\em interaction algebra\/},
which is the algebra $\cJ$ generated by
$\cJ_1=\vspan\{I,J_1,J_2,\ldots\}$, $I$ denoting the identity.  Thus,
elements of $\cJ$ are linear combinations of products of operators in
$\cJ_1$.  The linearly closed set $\cJ_1$ consists of the operators of
{\em degree\/} (at most) one.  Next define $\cJ_d=\cJ_1^d$, the linear
span of products of $d$ or less operators in $\cJ_1$. These are the
operators of {\em degree\/} $d$.  $\cJ_1$ is well-defined in the sense
that it is independent of the choice of operators $B_i$, provided that
they are linearly independent.  Because the interaction Hamiltonian is
Hermitian, $\cJ_1$ is $^\dagger$-closed, implying that $\cJ_d$ and
$\cJ$ are $^\dagger$-closed.

This general formalism easily translates to qubits and classical
error-correction.  If $S$ consists of $n$ qubits, then $N=2^n$, and a
{\em linear} interaction satisfies that each $J_i$ only involves Pauli
operators $\slb{\sigma}{k}_u$ acting on one qubit. Here $k$ is a qubit
label, $k\in\{1,\ldots,n\}$, and $u$ is one of $x$, $y$ or $z$.  {\em
Collective} linear interactions involve global operators $J_u =
\sum_k\slb{\sigma}{k}_u$, corresponding to a situation where a single
environment couples symmetrically to all
qubits~\cite{zanardi:qc1998a}.  In most discussions of quantum error
correction, however, the interaction is assumed to be {\em
independent}, meaning that each qubit interacts with its own
environment~\cite{knill:qc1995e}.  Independent interactions are
linear.  For linear interactions with qubits, $\cJ_d$ consists of
linear combinations of products of at most $d$ Pauli operators. It is
worth stressing that the Lie algebra generated by $\cJ_1$ usually
provides little information about higher-order errors.  For example,
the Lie algebra generated by the linear interactions contains only
linear interactions, while the effect of an environment coupled
linearly can include any other higher-order operator and is not
restricted to the unitary group generated by the Lie algebra.  The
{\em classical} error-correction problem is obtained by letting
$\cJ_1$ consist of all linear combinations of products of Pauli
operators with at most one $\sigma_x$ or $\sigma_y$ and arbitrarily
many $\sigma_z$'s. Thus, the interaction is completely (phase)
decohering, with first-order bit-flip noise.  We call this the {\em
classical interaction\/}.

\phead{Minimum Distance and Error Correction.} Noise for $S$ can now
be analyzed purely in terms of $\cJ_1$ and $\lambda$. By
straightforward generalization of the definitions for qubits and
independent interactions, we can define a minimum distance $d$ quantum
code for $S$ and $\cJ_1$ as a code that {\em
detects\/}~\cite{bennett:qc1996a} all errors in $\cJ_{d-1}$.  Recall
that a (quantum) code of $S$ is a subspace $\cC \subset \cS$, which
can be defined through the associated projector $\Pi_{\cC}$.  Error
$E$ is detected by $\cC$ if the following protocol works: 1. Prepare a
state $\ket{\psi}$ in $\cC$. 2. Allow error $E$ to occur, so that the
new state is $E\ket{\psi}$. 3. Make a measurement to detect whether
the state is in $\cC$ or in the orthogonal complement; the outcome is
either $\Pi_{\cC}E\ket{\psi}$ or $(I-\Pi_{\cC})E\ket{\psi}$. 4. Accept
the state in the former case and reject it otherwise.  The protocol is
correct if accepted states are proportional to the initial state, {\em
i.e.}, formally,
\begin{eqnarray}
\Pi_{\cC} E\,\Pi_{\cC} &=& \alpha_E \,\Pi_{\cC} \:,
\end{eqnarray}
for scalars $\alpha_E$.

In many cases we only need to preserve {\em classical} information.
We define a code to have minimum c-distance $d$ if for all $E,F\in
\cJ_{d-1}$, $\Pi_{\cC} E\,\Pi_{\cC}$ and $\Pi_{\cC} F\,\Pi_{\cC}$
commute. This implies that a basis of $\cC$ exists, such that the
above protocol is correct when restricted to basis elements.  We will
use the term {\em c-code\/} to denote a code intended only for
transmission of classical information in some basis. The notion of
error detection can be extended to c-codes if a transmission basis is
explicitly provided. We say that the c-code $\cC$ with orthonormal
basis $\ket{c_1},\ket{c_2},\ldots$ {\em detects\/} $E$ if $\Pi_{\cC}
E\,\Pi_{\cC}$ is diagonal when expressed in this basis, {\em
i.e.}~$\bra{c_i}E\ket{c_j} = \alpha_{i}\delta_{i,j}$.

An {\em $e$-error-correcting code\/} permits correction of all errors
in $\cJ_e$, which means that an initial state in the code can be
recovered by some fixed quantum operation after an error in $\cJ_e$ has
occurred.  Minimum distance is related to error correction in the
usual way.

\begin{Theorem}
A minimum (c-)distance $2e+1$ code is an $e$-error-correcting (c-)code. 
\end{Theorem}

\proof Recall the necessary and sufficient conditions for a code $\cC$
to permit correction of the errors in
$\cJ_e$~\cite{knill:qc1995e,nielsen:qc1998b}: $\cC$ detects the
operators in $\cJ_e^\dagger \cJ_e$.  This condition is also correct
for c-codes with a transmission basis. Since $\cJ_e^\dagger \cJ_e
\subset \cJ_{2e}$, the result follows.  \qed

\phead{Error Bounds.} To make the analysis based on minimum distance and 
$e$-error correction useful, it is necessary to show that 
$e$-error-correcting codes protect information well. We give a quantitative 
relationship for the worst-case error as a function of time $t$.

\begin{Theorem}
The error amplitude of information protected in an $e$-error-correcting
(c-)code is at most $(\lambda t)^{e+1}/(e+1)!$.
\label{thm:ebound}
\end{Theorem}

We defer the proof until after error-correcting codes have been
characterized as subsystems. Note that independence from the internal
Hamiltonian of the environment implies that even if the latter is
subject to arbitrary, adversarial manipulation, the error-correcting
code still effectively protects information on a time scale of
$O(1/\lambda)$.

\phead{Existence of Large Codes.}  A goal of constructing good
error-correcting codes is to maximize the dimension of minimum
(c-)distance $d$ codes. The greedy algorithm for constructing good
minimum-distance classical codes works well in the general case.  Let
$\{E_1=I,E_2,\ldots ,E_D\}$ be a basis of $\cJ_{d-1}$, with dimension
$D$, and let $\lceil x \rceil$ denote the least integer $\geq x$.

\begin{Theorem}
There exist codes of $S$ with minimum c-distance
$d$ of dimension at least $\left\lceil {N\over D}\right\rceil$.
\label{thm:ccode}
\end{Theorem}

\proof
Minimum c-distance is equivalent to the existence of an orthonormal basis 
$\ket{c_1},\ldots,\ket{c_k}$ of the c-code such that,
for each operator $E_l$ to be detected,
\begin{eqnarray}
\bra{c_i}E_l\ket{c_j} &=& \alpha_{i,l}\delta_{i,j} \:.
\label{eqn:bdet}
\end{eqnarray}
The proof greedily constructs such a basis. Let $\ket{c_1}$ be any 
state of $S$. Suppose that $\ket{c_1},\ldots,\ket{c_k}$ have been 
found, fulfilling (\ref{eqn:bdet}). Choose $\ket{c_{k+1}}$
orthogonal to the vectors $E_i\ket{c_j}$, $i=1,\ldots,D$, $j=1,\ldots,k$.
Such a state exists provided that $kD<N$. The new set of 
$\ket{c_i}$ satisfies (\ref{eqn:bdet}). Upon continuing until the set 
cannot be extended, a c-code of dimension at least $\left\lceil {N\over D}\right\rceil$ is found.
\qed


Our best general construction of good codes for {\em quantum} 
information is based on finding a subcode of a c-code.
\begin{Theorem}
There exist minimum distance $d$ codes of $S$ of dimension at least
$\left\lceil {N\over D}\right\rceil{1\over D+1}$.
\label{thm:qcode}
\end{Theorem}

\proof Let $\cC$ be a c-code of $S$ of dimension at least $\lceil
{N\over D}\rceil$ with basis $\ket{c_i}$ satisfying (\ref{eqn:bdet}).
Let $Y$ be the set of indices of the basis vectors. To construct a
large quantum code, we partition $Y$ into subsets $Y_i$ and seek
non-negative coefficients $\beta_{i,j}$, satisfying $\sum_{j\in
Y_i}\beta_{i,j}=1$. Let $\ket{q_i}=\sum_{j\in
Y_i}\sqrt{\beta_{i,j}}\ket{c_j}$. Then the orthonormal vectors
$\ket{q_i}$ span the desired code provided
\begin{eqnarray}
\bra{q_i}E_l\ket{q_j} &=& \gamma_l\,\delta_{ij}\:,\;\;\;\;\;\forall \:i,j,l\:.
\end{eqnarray}
Compute $\gamma_{l,i} = \sum_{j\in Y_i}\beta_{i,j}\alpha_{j,l}$, the
$\alpha_{j,l}$'s being given in (\ref{eqn:bdet}).  We need the
$\gamma_{l,i}$ to be independent of $i$. This problem can be cast in
terms of a convex sets problem. We need to find as many disjoint
subsets of the set of vectors $\vec{\alpha}_j = \{\alpha_{j,l}\}_l$
with the property that their convex closures have a common
intersection. Since $\cJ_{d-1}$ is $^\dagger$-closed,
the $\vec{\alpha}_j$ live in a subspace of real dimension $D$.
By invoking a generalization of Radon's
Theorem~\cite{tverberg:qc1964a}, a necessary condition for the
existence of at least $r$ such sets is $r(D+1)-D\leq \lceil
N/D\rceil$. The result follows. \qed

While the result of Thm.~\ref{thm:qcode} is fully general, the proof
does not yield a straightforward constructive method. Also, the lower
bound on the existence of good codes is sub-optimal for various
systems of interest, including qubits with linear interactions, where
the achieved rate is well below the best lower bounds
known~\cite{calderbank:qc1995a}.  According to Thms.~\ref{thm:ccode}
and~\ref{thm:qcode}, the above bounds for minimum distance $d$ codes
are found to be less favorable for independent than for collective
interactions, both for classical and for quantum information. This
reflects the lower level of complexity of the error process generated
by collective interactions.

\phead{Subsystems.}
If a system consists of a number of qubits, the
obvious subsystems are the qubits. If the system consists of a number
of photon modes, each mode is a subsystem.  However, in order to use
these modes as qubits, one could choose the two polarization states for 
a single photon in a mode as the computational basis. The
relevant system is then the subspace where each mode is occupied by
exactly one photon, and it is in this subspace that we can identify the
qubit subsystems. In both examples, subsystems appear as factors (in the 
tensor product sense) of subspaces of a larger state space. 
To avoid working with explicit bases and states, it is convenient to
resort to a general algebraic definition. We shall characterize a
subsystem of $S$ in terms of a subalgebra of operators acting on $\cS$
together with an irrep of the subalgebra. This is motivated by the
following fundamental result from the representation theory of
$^\dagger$-closed operator algebras~\cite{curtis:1962a}.

\begin{Theorem}
Let $\cA$ be a $^\dagger$-closed algebra of operators on $\cS$, 
including the identity. Then $\cS$ is isomorphic to a direct sum,
\begin{eqnarray}
\cS&\sim&\sum_i{\cC_i\tensor\cZ_i} \:,
\label{eqn:iso}
\end{eqnarray}
in such a way that in the representation on the righthandside, 
$\cA = \sum_i \text{Mat}(\cC_i)\tensor \slb{I}{Z_i}$ and the commutant 
of $\cA$ is given by 
$Z(\cA) = \sum_i \slb{I}{C_i}\tensor \text{Mat}(\cZ_i)$.
\label{thm:brep}
\end{Theorem}

Here, $\mbox{Mat}(\cH)$ denotes the set of all linear operators from 
$\cH$ to itself, while the commutant $Z(\cA)$ is the space of all operators 
commuting with $\cA$. Formally, each factor $Z_i$ $(C_i)$ in 
Thm.~\ref{thm:brep} defines a {\em subsystem} of $S$ with associated state
space $\cZ_i$ $(\cC_i)$. As a result of the theorem, subsystems are 
naturally definable in terms of either algebras or their commutants.

\phead{Noiseless Subsystems.} Consider the interaction algebra $\cJ$
associated with (\ref{eqn:j}). Since $\cJ$ is $^\dagger$-closed, the
representation of Thm.~\ref{thm:brep} applies. For each subsystem 
$Z_i$, states in $\cZ_i$ are completely immune to the interaction, as the
interaction operators only act on the co-factor $\cC_i$. Thus, $Z_i$
is a {\em noiseless subsystem i.e.\/}, a subsystem where information
is intrinsically stabilized against the effects of the noise.
As remarked above, a trivial example occurs if we are given two qubits, 
where only the second one is susceptible to noise. In this case, 
information in the first qubit is canonically maintained with no need 
for corrective action. Noiseless (or decoherence-free) 
subspaces~\cite{zanardi:qc1998a,lidar:qc1998a} can be recognized as special 
cases of the general decomposition (\ref{eqn:iso}) for appropriate 
interaction algebras $\cJ$. However, relevant situations can be devised, 
where noiseless subsystems exist in the absence of noiseless subspaces.

\phead{Example.}  
Let us consider three qubits $A,B,C$ with collective linear interactions. 
The interactions are the generators for spatial rotations. As pointed out 
in~\cite{zanardi:qc1998a}, no state of three qubits is invariant under 
spatial rotations, the minimal implementation of a noiseless subspace 
requiring $n=4$ qubits. 
However, the state space decomposes into one spin-$3\over 2$ and two
spin-$1\over 2$ irreducible subspaces. The two spin-$1\over 2$ components
together are representable as the product of two two-state spaces as
in Thm.~\ref{thm:brep}, with $\cJ$ acting only on the first. Thus the
second one is a noiseless subsystem. Another method of finding this
subsystem is to observe that the commutant $Z(\cJ)$ is non-trivial. In
particular, it includes the scalars under spatial rotations,
\begin{eqnarray}
s_1 &=& \slb{\sigma}{A}_x\slb{\sigma}{B}_x +
\slb{\sigma}{A}_y\slb{\sigma}{B}_y +
\slb{\sigma}{A}_z\slb{\sigma}{B}_z \:, \\
s_2 &=& \slb{\sigma}{A}_x\slb{\sigma}{C}_x +
\slb{\sigma}{A}_y\slb{\sigma}{C}_y +
\slb{\sigma}{A}_z\slb{\sigma}{C}_z \:, 
\end{eqnarray}
which are generating observables for the noiseless subsystem. Equivalently, 
the latter is seen to support one of the irreps of the algebra generated 
by the scalars.

\phead{Error-correcting Codes as Subsystems.} The traditional view of
error-correcting codes involves encoding the information and
correcting errors after the information carriers are transmitted
through a noisy channel. The concept of noiseless subsystem shows
that, for the purposes of information maintenance, it is not
necessary to correct errors, insofar as they affect components
independent of the system where information is stored.  In general, we
wish to protect the information against all errors in $\cJ_e$ for some
reasonably large $e$.  Since a subsystem unaffected by the operators
in $\cJ_e$ is automatically noiseless, but in most cases of interest
noiseless subsystems do not exist, it is necessary to take an active
role in maintaining information. Rather than using error correction to
restore the overall state of the system {\em after} errors happened,
we propose to use a quantum operation {\em before} the latter occur,
in such a way that the net effect of the quantum operation followed by
errors in $\cJ_e$ assures preservation of the information in a
subsystem.  A quantum operation is described by a family
$\cA=\{A_i\}_i$ of linear operators acting on $\cS$, evolving the
system density operator as $\rho$ $\mapsto$ $\sum_iA_i\rho
A_i^\dagger$.  The combined action of the quantum operation, followed
by errors in $\cJ_e$, is represented by the product of an operator $E
\in \cJ_e$ and one of the operators $A_i \in \cA$. Consequently, a
state of a noiseless subsystem of the $^\dagger$-closed algebra
generated by $\cJ_e\cA$ is preserved in this process.

\begin{Theorem}
Every $e$-error-correcting code arises as a noiseless subsystem of
$\cJ_e\cA$ for some $\cA$ with the property that
$I\in\vspan(\cA^\dagger\cA)$. Conversely, every noiseless subsystem of
$\cJ_e\cA$ with $\cA$ satisfying the above condition corresponds to an
$e$-error-correcting code.
\label{thm:ec=nl}
\end{Theorem}

\proof The fact that error-correcting codes yield such noiseless
subsystems follows from Thm.~III.5 of~\cite{knill:qc1995e} by letting
$\cA$ consist of operators that return the state of the error system
to the state $\ket{\cE(0)}$ (in the notation
of~\cite{knill:qc1995e}). Conversely, the condition 
$I\in\vspan(\cA^\dagger\cA)$ ensures the existence of a
quantum operation whose operators are in $\cA$.
Thus, the process suggested earlier protects the information against errors 
in $\cJ_e$. Because of the necessity of the conditions for error-correcting 
codes~\cite{knill:qc1995e}, there exists an associated
$e$-error-correcting code in the usual sense.  \qed

As a consequence, noiseless subsystems are infinite-distance
quantum error-correcting codes.

\phead{Error Analysis.} We are now ready to give a proof of 
Thm.~\ref{thm:ebound} based on viewing error-correcting codes as 
subsystems protected by an initial quantum operation $\cA$.

\proofof{~\ref{thm:ebound}} By purifying the
environment~\cite{schumacher:qc1996a}, we can assume that the
environment's initial state is $\kets{\psi}{B}$.  The
initial state of the system has the intended state in the subsystem
associated with the error-correcting code.  Again, by purifying and
by adding the reference system to $S$, we can assume that the state
is given by $\kets{\phi_0}{S}$. The quantum operation $\cA$ can be 
assumed to arise from a unitary evolution $U$ applied to 
$\kets{\phi_0}{S}\kets{0}{A}$, where $A$ is an ancillary system. 
Let $\ket{\phi} = U\kets{\phi_0}{S}\kets{0}{A}$ and consider
the subsequent interaction with the environment over time $t$.  By
slicing the interaction time into intervals of duration $t/n$, the 
overall evolution up to time $t$ can be written as ($\hbar=1$)
\begin{eqnarray}
&\lim_{n\rightarrow\infty}\prod_{k=1}^n
\slb{\delta U}{S}_k \slb{\delta U}{B}_k\ket{\phi}
  \kets{\psi}{B}\:,&
\end{eqnarray}
where $\slb{\delta U}{S}_k$, $\slb{\delta U}{B}_k$ denote the unitary
evolutions during the $k$'th interval due to $J$ and to the
environment's internal Hamiltonian respectively.  It suffices to
consider a first-order expansion $\slb{\delta U}{S}_k = I-iJ(t/n) +
O((t/n)^2)$.  The elements contributing noise all involve at least
$e+1$ factors of $J$. By distributing some of the sums $I-iJt/n$
starting at the first time interval, the expression inside the limit
can be thought of as a sum over the branches of a binary tree of
products of operators associated with the edges and nodes of the
tree~\cite{liu:1968}.  The root node is labeled $\slb{\delta U}{B}_1,$
and its two edges by $I$ and $-iJt/n$ respectively. The two children
are labeled by $\slb{\delta U}{B}_2$, their descendant edges by $I$
and $-iJt/n$ and so on. We choose to terminate a branch at a point
where there are $e+1$ factors of $-iJt/n$ on its path and label the
leaf with the remaining product of unitary operators. The total error
is estimated by summing the error amplitudes associated with the
products along each of these terminated branches. A counting argument
shows that there are ${n\choose e+1}$ such branches. Using $|CD|\leq
|C||D|$ and the fact that unitary operators preserve the amplitude,
the error of each such branch is bounded by $(\lambda
t/n)^{e+1}$. Thus, the error amplitude is at most ${n\choose
e+1}(\lambda t/n)^{e+1}\leq (\lambda t)^{e+1}/(e+1)!$.  \qed

\phead{Markovian Noise.} 
As mentioned above, in many cases involving infinite dimensional
environments the estimate of Thm.~\ref{thm:ebound} cannot be used 
without redefining $\lambda$. For instance, 
when the noise is to a good approximation Markovian, the evolution of the 
system density operator can be written as $\rho$ $\mapsto$
$\rho_t=\lim_{n\rightarrow\infty}\cL_{t/n}^n(\rho)$, where the 
superoperator $\cL_{t/n}$ takes the form 
$\cL_{t/n}(\rho) = \rho + (t/n)(\sum_i L_i\rho L_i^\dagger + V\rho + 
\rho V^\dagger)$ 
for a suitable choice of operators $L_i$ and $V$~\cite{alicki:1987}. 
Our techniques apply with $\cJ_1$ given by the 
linear span of $I$, these operators and their Hermitian transposes.
The bound of Thm.~\ref{thm:ebound} 
holds provided we replace error amplitude with error probability and 
redefine $\lambda$ as 
\begin{eqnarray}
\lambda &=& 2|V|+|L_1|^2+|L_2|^2+\ldots \:.
\label{eqn:lambdam}
\end{eqnarray}
The need to consider error probability rather than amplitude is due to
the statistical nature of Markovian noise.

\begin{Theorem}
The error probability of information protected in an $e$-error-correcting 
(c-)code subjected to Markovian noise is  
bounded by $(\lambda t)^{e+1}/(e+1)!$.
\label{thm:mbound}
\end{Theorem}

\proof Let $\rho$ be the state of the system after the protecting
quantum operation has been applied, and
$\rho_t=\lim_{n\rightarrow\infty}\cL_{t/n}^n(\rho)$.  If we write
$\rho_t = \rho_c + \rho_e$, where $\rho_c$ has no error in the
subsystem of interest, then the error probability is bounded by
$\trace(\sqrt{\rho_e^\dagger\rho_e})$.  The product of $n$
infinitesimal evolutions can be expanded as in the proof of
Thm.~\ref{thm:ebound}, but replacing unitary operators with
trace-preserving superoperators and omitting the environmental
contribution. The non-identity terms in the branches are
superoperators of the form $\rho''$ $\mapsto$ $(t/n)(\sum_i L_i\rho''
L_i^\dagger + V\rho'' + \rho'' V^\dagger)$, whose effect can be
bounded by the parameter $\lambda$ given in (\ref{eqn:lambdam}).  (Use
the fact that if we define $l_1(\rho) =
\trace(\sqrt{\rho^\dagger\rho})$, then $l_1(\rho U)\leq
l_1(\rho)|U|$).  \qed

In the past, error probability has been used in almost
all treatments of noise. Therefore, Thm.~\ref{thm:mbound} further
connects our formulation of error correction to the usual one.

\phead{Discrete Quantum Operations.} A picture of the situation
involving qubits coupled to independent environments, which is what has been
typically addressed by quantum error-correction theory, is that a known, 
somewhat noisy quantum operation is applied to each qubit. 
In this case, time does not play an explicit role. Instead, we are given 
quantum operations $\cL_i$, 
each expressible in the form $\cL_i(\rho)$
$=$ $(I+V_{i})\rho(I+V_{i}^\dagger)+ L_{i1}\rho L_{i1}^\dagger +
L_{i2}\rho L_{i2}^\dagger + \ldots$. The noise operation involves
applying each quantum operation to the system in some order.  To apply
our theory and bounds to this situation, define $\cJ_1$ as the span of
$I$, the operators $V_{i}$, $L_{ij}$ and their Hermitian transposes. 
The noise-strength parameter can be redefined as $\lambda
= \max_i(2|V_i| + |V_i|^2 + \sum_{j} |L_{ij}|^2)$, so that
Thm.~\ref{thm:mbound} applies for $t=1$. This gives estimates not
far from the standard ones in the case of qubits subject to depolarizing
noise~\cite{bennett:qc1996a}.

\phead{Time-Dependent Noise.}
Time-dependent noise can arise from the use of a rotating frame to
compensate for internal evolution of the system, or to compensate for
time-dependent control actions, such as the ones exploited in decoupling
schemes for open quantum systems~\cite{viola:qc1998a}. 
To adapt our theory to this situation, it suffices to maximize the 
expression of $\lambda$ over time and choose $\cJ_1$ as the span of 
all first-order operators occurring at various times.

\phead{Summary.} By suitably incorporating the description of the error
generation process within a general algebraic setting, we showed how to 
reformulate quantum error correction without restricting the statistical
properties of the environmental noise. 
The existence of large codes was established for both classical and quantum 
information, opening the way to accurate quantum computations in the 
presence of arbitrary errors. 
In addition to substantially strengthening the power of quantum 
error-correction theory, our analysis points to the notion of a noiseless 
subsystem as an emerging unifying framework for quantum information 
protection. Full exploitation of the above concept should prove 
fruitful in the general context of quantum information processing.

\phead{Acknowledgments.} We thank Asher Peres for the three qubit
example of a noiseless subsystem.  E. K. and R. L. received support
from the Department of Energy, under contract W-7405-ENG-36, and from
the NSA.  
L. V. was supported in part by DARPA/ARO under the QUIC
initiative.


\begin{thebibliography}{10}

\bibitem{shor:qc1996a}
P.~W. Shor.
\newblock Fault-tolerant quantum computation.
\newblock In {\em Proceedings of the Symposium on the Foundations of Computer
  Science}, pages 56--65, Los Alamitos, California, 1996. IEEE press.
\newblock {\tt quant-ph/9605011}.

\bibitem{aharonov:qc1996a}
D.~Aharonov and M.~Ben-Or.
\newblock Fault-tolerant quantum computation with constant error.
\newblock In {\em Proceedings of the 29'th Annual ACM Symposium on the Theory
  of Computing}, pages 176--188, New York, 1996. ACM Press.
\newblock {\tt quant-ph/9611025}.

\bibitem{kitaev:qc1997a}
A.~Yu. Kitaev.
\newblock Quantum computations: algorithms and error correction.
\newblock {\em Uspekhi Mat. Nauk.}, 52:53--112, 1997.

\bibitem{knill:qc1998a}
E.~Knill, R.~Laflamme, and W.~H. Zurek.
\newblock Resilient quantum computation.
\newblock {\em Science}, 279:342--345, 1998.

\bibitem{shor:qc1995b}
P.~W. Shor.
\newblock Scheme for reducing decoherence in quantum computer memory.
\newblock {\em Physical Review A}, 52:R2493--R2496, 1995.

\bibitem{steane:qc1995a}
A.~Steane.
\newblock Multiple particle interference and quantum error correction.
\newblock {\em Proceedings of the Royal Society of London A}, 452:2551--2577,
  1996.

\bibitem{knill:qc1995e}
E.~Knill and R.~Laflamme.
\newblock A theory of quantum error correcting codes.
\newblock {\em Physical Review A}, 55:900--911, 1997.

\bibitem{zanardi:qc1998a}
P.~Zanardi and M.~Rasetti.
\newblock Noiseless Quantum Codes
\newblock  {\em Physical Review Letters}, 79:3306--3309, 1998.
\newblock P.~Zanardi and F.~Rossi.
\newblock Quantum information in semiconductors: Noiseless encoding in a
  quantum-dot array.
\newblock {\em Physical Review Letters}, 81:4752--4755, 1998.

\bibitem{lidar:qc1998a}
D.~A. Lidar, I.~L. Chuang, and K.~B. Whaley.
\newblock Decoherence-free subspaces for quantum computation.
\newblock {\em Physical Review Letters}, 81:2594--2597, 1998.

\bibitem{bennett:qc1996a}
C.~H. Bennett, D.~P. DiVincenzo, J.~A. Smolin, and W.~K. Wootters.
\newblock Mixed state entanglement and quantum error-correcting codes.
\newblock {\em Physical Review A}, 54:3824--3851, 1996.

\bibitem{nielsen:qc1998b}
M.~A. Nielsen, C.~M. Caves, B.~Schumacher, and H.~Barnum.
\newblock Information-theoretic approach to quantum error-correction and
  reversible measurement.
\newblock {\em Proceedings of the Royal Society of London A}, 454:277--304,
  1998.

\bibitem{tverberg:qc1964a}
H.~Tverberg.
\newblock A generalization of Radon's theorem.
\newblock {\em J. London Math. Soc.}, 41:123--128, 1966.

\bibitem{calderbank:qc1995a}
A.~R. Calderbank and P.~W. Shor.
\newblock Good quantum error-correcting codes exist.
\newblock {\em Physical Review A}, 54:1098--1105, 1996.

\bibitem{curtis:1962a}
C.~W. Curtis and I.~Reiner.
\newblock {\em Representation Theory of Finite Groups and Associative
  Algebras}.
\newblock Interscience Publishers, Wiley\&Sons, 1962.

\bibitem{schumacher:qc1996a}
B.~Schumacher.
\newblock Sending entanglement through noisy quantum channels.
\newblock {\em Physical Review A}, 54:2614--2628, 1996.

\bibitem{liu:1968} C.~L.~Liu.
\newblock {\em Introduction to Combinatorial Mathematics}
\newblock McGraw-Hill, New York, 1968.

\bibitem{alicki:1987} 
R.~Alicki and K.~Lendi.
\newblock {\em Quantum Dynamical Semigroups and Applications}.
\newblock Springer-Verlag, Berlin, 1987.

\bibitem{viola:qc1998a}
L.~Viola and S.~Lloyd.
\newblock Dynamical suppression of decoherence in two-state quantum systems.
\newblock {\em Physical Review A}, 58:2733--2744, 1998.
\newblock L.~Viola, E.~Knill, and S.~Lloyd.
\newblock Dynamical decoupling of open quantum systems.
\newblock {\em Physical Review Letters}, 82:2417--2421, 1999.

\end{thebibliography}
\end{document}